\documentclass[lettersize,journal]{IEEEtran}
\usepackage{amsmath,amsfonts,amssymb}
\usepackage{algorithm}
\usepackage{algpseudocode}
\usepackage{array}
\usepackage{tabularx}
\usepackage{booktabs}
\usepackage{enumitem}
\usepackage[caption=false,font=normalsize,labelfont=sf,textfont=sf]{subfig}
\usepackage{textcomp}
\usepackage{stfloats}
\usepackage{url}
\usepackage{verbatim}
\usepackage{graphicx}
\usepackage{cite}
\usepackage{xcolor}
\usepackage[normalem]{ulem}
\usepackage{tcolorbox}
\tcbuselibrary{listings}
\usepackage{multirow}
\usepackage{listings}
\usepackage{xspace}
\usepackage{caption}
\usepackage{makecell}
\usepackage{hyperref}

\usepackage[frozencache]{minted}

\newcommand{\framework}{\textsc{PoCEvolve}\xspace}

\newcommand{\pocgen}{\textsc{PoCGen}\xspace}

\begin{document}

\title{\framework: Generating Proof-of-Concept Exploits from Security Patches with Vulnerability-Aware Prompt Evolution}

\author{Duc Manh Tran,
        Ratnadira Widyasari,
        Ivana Clairine Irsan,
        Huihui Huang,
        Ting Zhang\textsuperscript{*},
        Shar Lwin Khin,
        Ouh Eng Lieh,
        Hong Jin Kang,
        and David Lo%
\thanks{Duc Manh Tran and Hong Jin Kang are with The University of Sydney, Sydney, NSW, Australia.}
\thanks{Ratnadira Widyasari, Ivana Clairine Irsan, Huihui Huang, Shar Lwin Khin, Eng Lieh Ouh, and David Lo are with Singapore Management University, Singapore.}%
\thanks{Ting Zhang is with Monash University, Australia. \textsuperscript{*}Corresponding author.}%
}

\maketitle

\begin{abstract}
Ideally, the detailed information about a vulnerability should be made available together with the fixing commit.
In practice, however, such details often become available only long after the commit, even when a CVE has already been published.
During this window, the patch is already public, so attackers can reverse-engineer it, yet defenders lack the details needed to assess exposure, prioritize, and validate the fix. 
Executable evidence, such as a proof-of-concept (PoC) exploit, could fill this gap. 
Prior work has automated PoC generation, but the state-of-the-art approach, \pocgen, assumes that a detailed vulnerability report is already available, which is precisely what is missing during this window. 
In this paper, we first present an empirical study quantifying the long delay between the fixing commit and the availability of a detailed vulnerability report.
We then introduce \framework, a vulnerability-aware prompt-evolution framework that generates PoCs directly from vulnerability-fixing commits.
Given a vulnerability-fixing commit, \framework synthesizes a corresponding PoC exploit.
To learn from unsuccessful generation attempts, \framework assesses the usefulness of different dimensions of vulnerability-related context, including the inferred vulnerable API and code-coverage information.
These assessments guide prompt evolution towards more effective exploit-generation prompts. 
We evaluate \framework on SecBench.VFC.js, where \framework achieves a PoC generation success rate of 58.4\%, corresponding to relative improvements of 20.7\% over \pocgen and 200.0\% over the LLM baseline with GPT-4o-mini.
With a recent model, Qwen3.7-Plus, \framework achieves a higher success rate of 85.3\%.
When detailed vulnerability reports are available, \framework achieves a success rate of 71.7\%, improving over \pocgen by 11.1\%.
Across the proposed-setting configurations, the average cost per evaluated vulnerability ranges from \$0.0292 for \framework (LLM) with GPT-4o-mini to \$0.2492 for the full \framework with Qwen3.7-Plus.
\end{abstract}


\section{Introduction}
When a security vulnerability is discovered, it is reported to the software development team and a CVE Numbering Authority (CNA).
The vulnerability is assigned a CVE identifier and, after the relevant evidence is collected and the vulnerability is confirmed, the CVE is publicly disclosed~\cite{cymulate_cve}.
Crucially, the security patch and this public disclosure rarely occur at the same time.
Vulnerabilities are often patched before they are publicly disclosed~\cite{li2017large, wang2019detecting, zhou2021finding, dong2025lies}, and the delay between the two can be substantial~\cite{li2017large}.
This gap creates a significant window in which attackers can infer the vulnerability from the public patch, while downstream users may remain unaware that their deployed systems are affected and require urgent updating~\cite{10.1109/SP.2008.17}.
During this interval, defenders need executable evidence to determine whether their systems are affected and to validate that a patch removes the vulnerable behavior.
However, such evidence is often unavailable because a detailed vulnerability report is not yet available.

To assess how common this patch-only setting is in practice, we conduct an empirical study of patching practices across 26,803 CVEs.
Our analysis shows that security fixes usually precede public disclosure: 97.1\% of the CVEs in our dataset were patched before they were publicly disclosed, with a median patch-to-disclosure gap of 18 days and a mean gap of 147 days.
Thus, for many vulnerabilities, the patch is available before defenders have access to a public report or advisory.
Moreover, over 70\% of CVE patches contain no test files, making it challenging for downstream users to confirm the vulnerable behavior or validate the patch. 
These findings motivate the problem we address: generating PoC exploits directly from vulnerability-fixing commits, before a detailed vulnerability report is available.
Automated PoC generation is a natural way to supply this missing evidence.

Existing techniques generate PoCs through symbolic execution~\cite{kang2023scaling, marques2025automated} or fuzzing~\cite{cassel2025nodemedic} to discover inputs that trigger vulnerabilities. 
More recently, Simsek et al.~\cite{simsek2025pocgen} proposed \pocgen, which combines LLMs with program analysis to generate exploits from vulnerability reports.
\pocgen outperforms prior approaches~\cite{cassel2025nodemedic,kang2023scaling,marques2025automated}, and is the state-of-the art approach in automated PoC generation, which makes it our closest point of comparison.
\pocgen, however, assumes that a vulnerability report is available.
It does not operate directly from the vulnerability-fixing commit alone, which is often the only information available after the vulnerability is fixed and before security engineers have analyzed the vulnerability. 

Generating PoC exploits directly from vulnerability-fixing commits is challenging for three reasons. 
First, unlike detailed vulnerability reports, commit messages and diffs do not directly disclose critical information, such as the vulnerability's root cause or affected API~\cite{yang2025code, song2025not}. 
Second, vulnerability-fixing commits do not necessarily show how the vulnerability can be triggered; constructing a valid PoC exploit requires understanding the broader repository context, including API usage patterns and runtime assumptions. 
Third, vulnerability-triggering conditions are heterogeneous and sensitive: small differences in payload length, structure, or environment setup can cause an exploit to fail. 

To address these challenges, we formulate a new PoC generation setting in which only a vulnerability-fixing commit is provided without a vulnerability report, while the vulnerable repository and package context remain accessible for analysis.
In this setting we present \framework, a framework for generating PoC exploits from security patches.
First, \framework reconstructs the missing link between the underlying vulnerability and how it can be triggered: it extracts security-relevant information from the commit, such as the vulnerability type and affected APIs, and uses it to reconstruct a vulnerability report, which is passed to an exploit generator that produces multiple candidate PoC exploits.
Second, to cope with the small differences that make exploit synthesis brittle, \framework learns from failed attempts: it records the prompts behind failed candidates and scores each along several vulnerability-context dimensions, then generates feedback identifying the dimension along which the prompt should be revised.
This vulnerability-context-guided feedback drives the evolution of the exploit-generation prompt, allowing it to better target the observed failures.

On our SecBench.VFC.js benchmark of real-world vulnerabilities,
constructed from 190 vulnerabilities of SecBench.js by filtering for cases with available vulnerability-fixing commits,
we find that prompt evolution improves both evaluated Exploit Generators without access to vulnerability reports.
With GPT-4o-mini, \framework (LLM) improves direct LLM prompting from 19.5\% to 28.4\%, while the full \framework improves \pocgen from 48.4\% to 58.4\%.
With Qwen3.7-Plus, the corresponding success rates increase from 77.9\% to 85.3\% and from 64.2\% to 79.5\%, respectively. 
We also evaluate \framework when vulnerability reports are available. In this setting, \framework improves over \pocgen by 11.1\%, indicating that prompt evolution remains useful when more vulnerability information is available.
In the proposed setting, the average cost per evaluated vulnerability ranges from \$0.0292 to \$0.2492 across the evaluated model and Exploit Generator configurations.

The main contributions of this paper are as follows:

\begin{itemize}
    \item We formulate a new problem setting for generating PoC exploits directly from vulnerability-fixing commits and conduct an empirical study showing that this patch-only setting is prevalent in practice.
    \item We propose \framework, which leverages feedback from vulnerability-related contextual dimensions to iteratively improve exploit-generation prompts. \framework features a vulnerability-context-guided Prompt Evolver that evaluates failed prompts and provides feedback for prompt refinement.
    \item We conduct an extensive evaluation of \framework in both the proposed and report-guided settings, demonstrating its effectiveness across different underlying LLMs and input conditions, as well as its practical monetary and runtime costs.
\end{itemize}

The remainder of this paper is organized as follows. 
Section~\ref{sec:background} provides background on exploit generation and presents our empirical study of CVE patch practices,
Section~\ref{sec:motivation} illustrates the motivation of this work through an example,
Section~\ref{sec:problem} presents the problem statement, 
Section~\ref{sec:approach} describes the \framework approach, 
Section~\ref{sec:settings} details our evaluation setup, 
Section~\ref{sec:results} reports the experimental results, 
Section~\ref{sec:discussion} provides a qualitative analysis, failure analysis, and discusses threats to validity. 
Section~\ref{sec:related-work} reviews related work, 
and finally, Section~\ref{sec:conclusion} concludes the paper.

\section{Background}
\label{sec:background}

\subsection{Empirical Study on Characteristics of CVE Patch Practices}
\label{sec:empirical_study}

We collected a snapshot of 359,627 CVE records from the official CVE List repository~\cite{cveproject_cvelistv5} on 22/06/2026, covering CVEs published since 1999.
Among them, 31,935 CVEs reference a GitHub patch link.
Using the GitHub API, we successfully retrieved the patches for 26,803 CVEs and analyzed this subset of CVEs.
Our analysis reveals three findings that motivate this work.

\paragraph*{Patch-to-disclosure gap}
We use a CVE's publication date as its public disclosure timestamp, i.e., the point at which the vulnerability first becomes publicly known.
For 97.1\% of CVEs, the fix was committed before this disclosure, with a median gap of 18 days and a mean of 147 days between the earliest patch commit and disclosure.
Since a detailed vulnerability report is often released only after the CVE is published, the window during which a fix is public but no detailed report is available is at least this long.
Public patch diffs can reveal the underlying vulnerability, so attackers may develop exploits before defenders receive sufficient information~\cite{10.1109/SP.2008.17}.
This motivates techniques that can analyze patches and generate executable PoCs as soon as a fix becomes available.

\paragraph*{JavaScript ecosystem dominance}
The share of JavaScript CVEs increased from below 10\% before 2018 to 19.5\% in 2025 and 23.3\% in 2026, making it the largest language ecosystem in both years.
This trend indicates that the JavaScript ecosystem has become an increasingly prominent source of vulnerabilities and is likely to attract further attacker attention as its adoption continues to grow.
We therefore focus on JavaScript vulnerabilities as a timely and security-critical target for automated PoC generation.

\paragraph*{Absence of test cases in patches}
We find that 70.2\% of CVE patches contain no test files.
We identify test files by checking whether the changed file path contains \texttt{test} or \texttt{tests}.
Without tests, patches provide no machine-executable specification of the vulnerable behavior or its triggering conditions.
This finding directly motivates our investigation of automated PoC generation as a complement to patch analysis.

\subsection{Proof-of-Concept Exploit Generation Approaches}

Automated PoC exploit generation aims to synthesize executable inputs that trigger software vulnerabilities. 
Prior work uses symbolic execution~\cite{kang2023scaling, marques2025automated}, fuzzing~\cite{cassel2025nodemedic}, and program analysis-assisted exploit synthesis. 
However, they struggle to reason about complex program behaviors.

Recently, LLM-based approaches have been shown to be effective.
\pocgen~\cite{simsek2025pocgen} is the state-of-the-art approach for automatically generating PoC exploits for real-world vulnerabilities. 
Given a vulnerability report, \pocgen combines LLMs with static program analysis to identify relevant vulnerability information and synthesize candidate exploits. 
Generated exploits are executed and validated in a controlled environment, where runtime feedback from a validation mechanism is iteratively used to refine failed exploit attempts. 
This feedback-driven refinement process enables robust exploit synthesis.
However, this refinement depends on incorporating manual feedback into the prompt for the LLM to successfully generate an exploit.
For example, if an exploit executes a shell command but fails to achieve the expected outcome, the validation mechanism in \pocgen appends a fixed, prewritten hint into its prompt: \texttt{"The reason for this might be that the command is not injected properly or escaped."} 
Each prewritten hint handles a predefined failure pattern.
This suggests the need for automatically adapting prompts based on their failures in generating an exploit that triggers the vulnerability.

\subsection{Self-Evolving Systems}

Self-evolving systems aim to iteratively improve generation quality by leveraging feedback from previous outputs. 
In the context of LLM-based systems, prior work has explored iterative refinement, reflection-based prompting, and prompt evolution techniques to improve reasoning and code generation performance~\cite{zhou2023large, pmlr-v235-fernando24a, madaan2023self, shinn2023reflexion, pryzant2023automatic, gou2024critic, yang2024large}. These approaches typically use execution feedback, error traces, or model-generated reflections to revise prompts and guide subsequent generations.

Genetic-Pareto Prompt Optimization (GEPA)~\cite{agrawal2026gepa}, a state-of-the-art prompt evolution technique, enables multi-objective prompt evolution. 
Instead of targeting a single objective, as is done with simple runtime feedback, GEPA jointly considers multiple dimensions.
After generating multiple candidate prompts, the prompts are evaluated using LLM-based reflection to analyze the failures and propose targeted improvements.
A Pareto-based selection strategy is then applied to retain prompts that achieve strong trade-offs across objectives, encouraging both effectiveness and diversity among evolved prompts.
Compared to prior prompt-evolution approaches, GEPA incorporates structured feedback and multi-objective optimization, enabling robust prompt refinement.

As \pocgen~\cite{simsek2025pocgen}, the state-of-the-art Exploit Generator, relies on a static prompt construction strategy, it suffers from the same limitations that prompt-evolution methods are designed to address.
However, general prompt-evolution methods do not specify how vulnerability-related context should be evaluated.
This motivates our design of task-specific criteria that provide vulnerability-specific optimization signals for PoC exploit generation.

\section{Motivating Example}
\label{sec:motivation}

\begin{listing}[!t]
\begin{minted}[
frame=lines,
framesep=2mm,
fontsize=\scriptsize
]{diff}
// lexer.js
-lexer.addRule(/"((?:[^"\\]+|\\.)*)("|$)/, ...)
+lexer.addRule(/"((?:\\.|[^"])*)($|")/, ...)

\end{minted}
\caption{Vulnerability-fixing commit for npm:dirty-json:20180213: the inner quantifier \texttt{[{\textasciicircum}"{\textbackslash}{\textbackslash}]+} (one-or-more) is replaced by \texttt{[{\textasciicircum}"]} (single character), eliminating the ambiguity that causes catastrophic backtracking when matching against quoted strings.}
\label{code:dirtyjson-vfc}
\end{listing}

\begin{listing}[t]
\begin{minted}[
frame=lines,
framesep=2mm,
fontsize=\scriptsize
]{javascript}
// dirty-json/lexer.js
lexer.addRule(/"((?:[^"\\]+|\\.)*)("|$)/, (lexeme, txt) => {
    return { type: LEX_QUOTE, value: stripslashes(txt) }
})
\end{minted}
\caption{Vulnerable code in \texttt{dirty-json} (npm:dirty-json:20180213).}
\label{code:dirtyjson-vuln}
\end{listing}

\begin{listing}[!t]
\begin{minted}[
frame=lines,
framesep=2mm,
fontsize=\scriptsize
]{javascript}
// Exploit Code 1
const lexer = require("dirty-json/lexer");
// input is missing backslash sequences
const input = '"'.repeat(30000) + '"';
const result = await lexer.lexString(input, (token) => { });

// Exploit Code 2
const lexer = require("dirty-json/lexer");
const maliciousInput = '"\\'.repeat(1000) + '"';
// missing a function call to `emit`
const result = await lexer.lexString(maliciousInput);

// Exploit Code 3
const lexer = require("dirty-json/lexer");
// insufficient length
let payload = '"\\'" + '".repeat(10000) + '"';
let tokens = [];
const emit = (token) => { tokens.push(token); };
lexer.lexString(payload, emit);
\end{minted}
\caption{Representative failed exploit candidates generated by \pocgen.}
\label{code:pocgen-generated-exploits}
\end{listing}

We conduct a preliminary experiment on the \texttt{dirty-json}\cite{dirtyjsonnpm} npm package, which suffered from a Regular Expression Denial of Service (ReDoS) vulnerability~\cite{dirtysnyk}.
The \texttt{dirty-json} package is designed to parse non-conforming JSON inputs.
Listing~\ref{code:dirtyjson-vfc} shows its vulnerability-fixing commit, titled \textit{``(probably) fixed REDOS issues,''} which replaces the inner quantifier \texttt{[{\textasciicircum}"{\textbackslash}{\textbackslash}]+} (one or more) with \texttt{[{\textasciicircum}"]} (single character), eliminating the cause of catastrophic backtracking.

From its vulnerability-fixing commit, the vulnerable code and the type of vulnerability, a ReDoS, are identified.
\pocgen generates multiple PoC candidates,
shown in Listing~\ref{code:pocgen-generated-exploits}.
Exploit Code~1 uses a sufficiently long input, but it fails to include quoted strings.
Exploit Code~2 constructs the payload using repeated escaped-character patterns, but omits the \texttt{emit} callback.
Exploit Code~3 includes the required \texttt{emit} callback and a partially correct input structure, but it is not long enough.

These failed attempts reveal the requirements for a successful PoC exploit, which must preserve three key elements: 1) a sufficiently large input, 2) an escaped-character pattern that exercises the vulnerable quoted-string regular expression, and 3) the \texttt{emit} callback required by the lexer.
\pocgen does not consistently preserve these critical elements as it refines the exploit and improves it along individual dimensions.
For example, \pocgen removes the \texttt{emit} callback when improving the payload structure.
This highlights a key limitation of runtime feedback alone, as partial progress is not recognized and can be discarded.

A successful exploit generated by \framework, shown in Listing~\ref{code:dirtyjson-correct-exploit}, combines the key characteristics from the failed attempts.
Although no individual attempt constitutes a valid exploit, the attempts collectively suggest a viable exploit strategy.

\begin{listing}[t]
\begin{minted}[
frame=lines,
framesep=2mm,
fontsize=\scriptsize,
breaklines
]{javascript}
const lexer = require("dirty-json/lexer");
// Escaped-character pattern with sufficient length
const payload = '\\"'.repeat(20000) + '\\"' + 'a'.repeat(30000);
const tokens = [];
// emit callback required by lexString()
const emit = (token) => { tokens.push(token); };
lexer.lexString(payload, emit);
\end{minted}
\caption{Correct exploit after learning from failed attempts.}
\label{code:dirtyjson-correct-exploit}
\end{listing}

\section{Problem Statement}
\label{sec:problem}

We formulate the task of automatically generating PoC exploits from vulnerability-fixing commits without ground-truth vulnerability reports.
Unlike prior work~\cite{simsek2025pocgen}, which assumes access to report-provided information such as the affected API and vulnerability type, our setting provides the vulnerability-fixing commit as the primary vulnerability-specific artifact. 
We assume that the commits made to open-source repositories may be monitored for changes that fix security issues. Prior work identifies such monitoring as a realistic capability for both adversaries and security practitioners~\cite{li2017large,sawadogo2022sspcatcher,nguyen2022vulcurator,zhou2022spi}.

Let \(C=(m,d)\) denote a vulnerability-fixing commit, where \(m\) is the commit message and \(d\) is the code diff that patches a vulnerable program \(P\).
Let \(R\) denote the accessible repository and package context, including the vulnerable source code, package metadata, documentation, and usage examples.
Given \(C\) and \(R\), but without a ground-truth vulnerability report, the goal is to synthesize a PoC exploit \(E\) through a generator \(f\):

\[
E = f(C,R) = f(m,d,R).
\]

\section{Approach}
\label{sec:approach}

\begin{figure}[!t]
    \centering
    \includegraphics[width=1\linewidth]{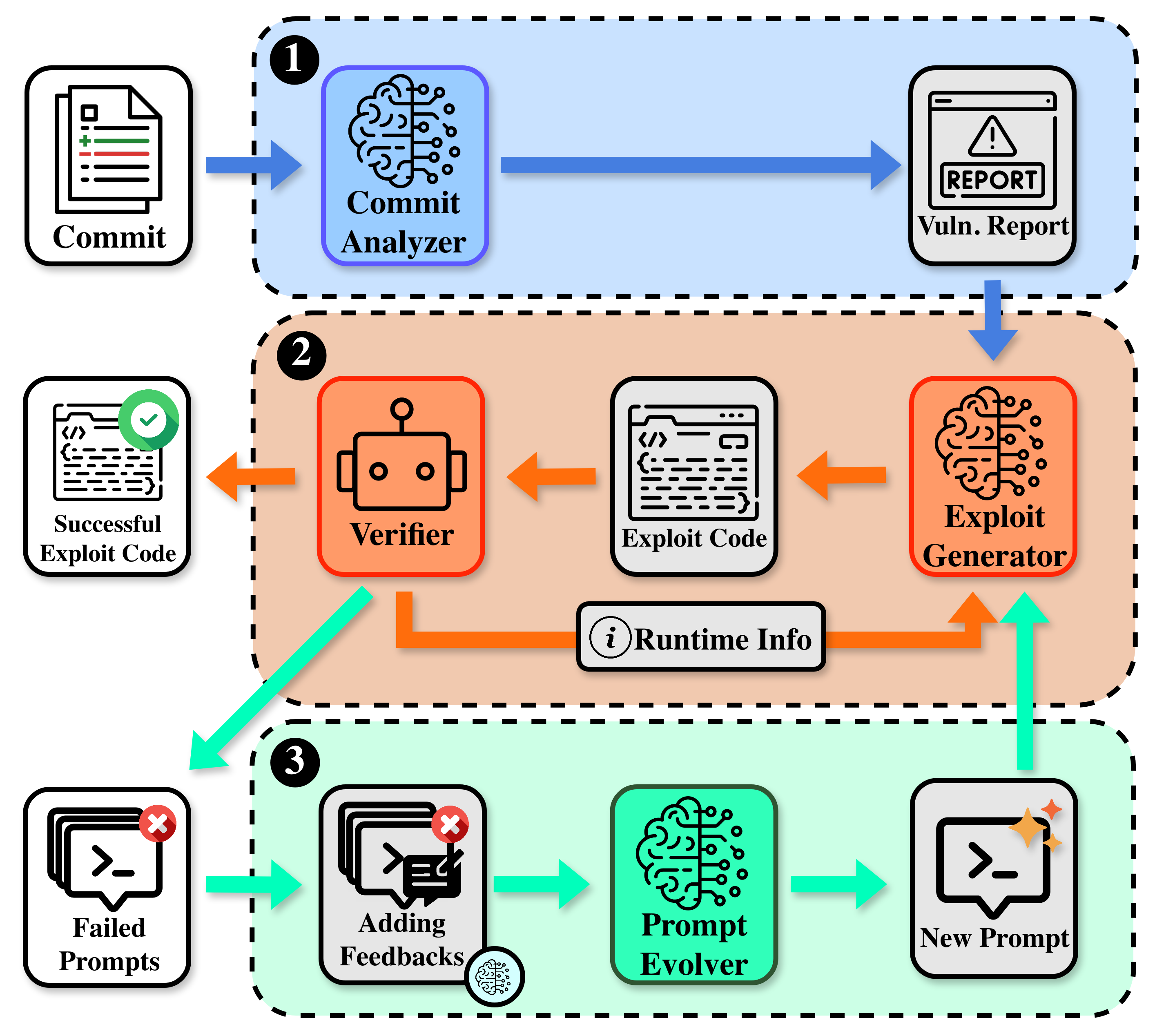}
    \caption{Overview of \framework. Given a vulnerability-fixing commit, the Commit Analyzer generates a structured vulnerability report, which guides the Exploit Generator in producing candidate PoC exploits. The Verifier evaluates each candidate and returns execution feedback. After repeated failures, the Prompt Evolver analyzes previous attempts and synthesizes an improved prompt for the next generation iteration.}
    \label{fig:overview}
\end{figure}

Figure~\ref{fig:overview} presents an overview of \framework, which operates in three phases. 
In Phase 1, given a vulnerability-fixing commit and accessible repository context, the Commit Analyzer reconstructs a vulnerability report containing the vulnerability type, vulnerable API, and a natural-language description of the vulnerability (Section~\ref{sec:analyzer}). No ground-truth vulnerability report is provided in this phase.
In Phase 2, the Exploit Generator uses this report to produce candidate PoC exploits (Section~\ref{sec:exploit-generator}), while the Verifier validates each candidate using runtime feedback and security-specific checks (Section~\ref{sec:Verifier}). 
If validation succeeds, \framework outputs the exploit code. Otherwise, after repeated unsuccessful attempts, Phase 3 invokes the Prompt Evolver, which analyzes the failed generation prompts, incorporates additional feedback, and synthesizes an improved prompt for the next generation attempt (Section~\ref{sec:evolver}). 
This process continues until a valid exploit is produced or the resource budget is exhausted.

\subsection{Vulnerability-Fixing Commit Analyzer}
\label{sec:analyzer}

The Commit Analyzer first employs an LLM to classify the vulnerability fixed by each commit into its vulnerability type. 
Following Simsek et al.~\cite{simsek2025pocgen}, we focus our evaluation on five vulnerability types commonly observed in JavaScript programs: \emph{Path Traversal}, \emph{Prototype Pollution}, \emph{Command Injection}, \emph{Code Injection}, and \emph{ReDoS}.
The analyzer then identifies vulnerability-related information from the package metadata, commit message, and code changes, including the vulnerable API, modified code regions, and relevant file locations. Finally, it generates a structured vulnerability description,
which later guides downstream exploit generation.
The analyzer prompt uses the package name, vulnerable version, commit message, and code changes as input, and constrains the LLM to return a JSON object containing the predicted vulnerability type, vulnerable API, and concise vulnerability description. 

\subsection{Exploit Generator}
\label{sec:exploit-generator}

After vulnerability analysis, the Exploit Generator produces candidate PoC exploits based on structured inputs provided by the analyzer. The generator takes as input (i) the vulnerability type, (ii) the vulnerable API, including function names, module imports, and file paths, (iii) the relevant code context or diff snippet, and (iv) the associated commit message. Together, these signals provide both the semantic and syntactic context required for constructing a targeted exploit.

The design of \framework is modular, allowing any capable system to serve as the Exploit Generator. 
In this work, we adopt \pocgen as the default generator due to its strong empirical performance and its ability to generalize across diverse vulnerability types. Prior work reports that \pocgen achieves a success rate of approximately 77\% on SecBench.js, significantly outperforming alternative approaches such as \textit{Explode.js} (50--60\%), which does not leverage LLMs, and \textit{AutoGPT}-based approaches (25--35\%), which suffer from high computational overhead and cost.

Beyond successful generations, \framework also explicitly collects failed generation attempts produced during refinement. These include user prompts that have been iteratively modified across the generator-verifier loop but still fail verification. 
Although labeled as unsuccessful by the verifier, these failed exploit candidates are informative, as they preserve rich contextual signals reflecting how the model attempted to repair earlier mistakes. This failure corpus is later leveraged by the Prompt Evolver to guide structured refinement and improve subsequent generations.

\subsection{Prompt Evolver}
\label{sec:evolver}

\begin{figure}
\centering
\includegraphics[width=1\linewidth]{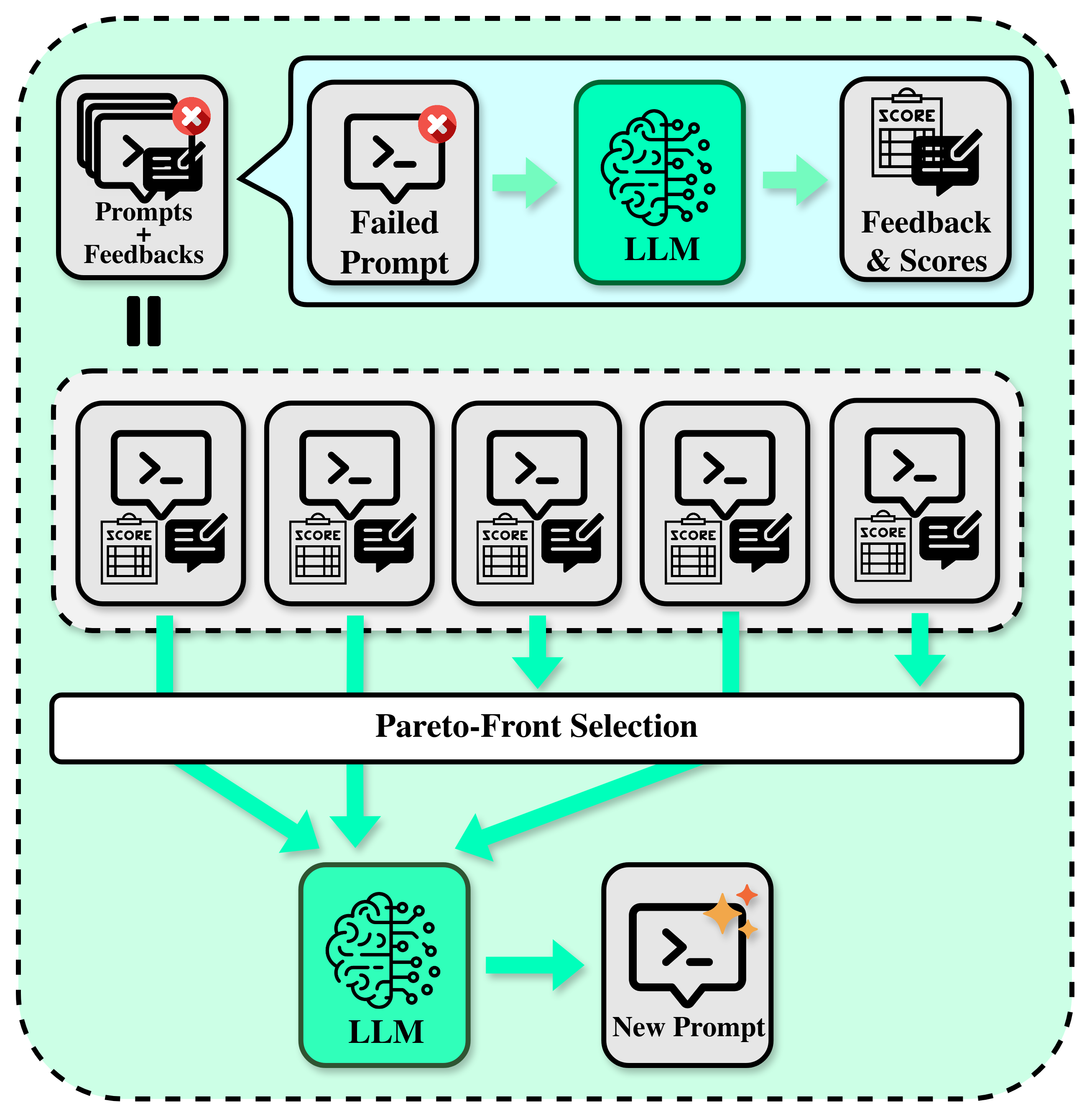}
\captionsetup{font=normalsize}
\caption{Overview of the Prompt Evolver. Starting from failed prompt candidates, the Prompt Evolver uses an LLM to assess the usefulness of vulnerability-related context, derives structured feedback, maintains a pool of prompt-feedback-score candidates, selects Pareto-front candidates, and synthesizes a new prompt for the next generation iteration.}
\label{fig:prompt-evolver}
\end{figure}

To improve on \pocgen's heuristic refiner, the Prompt Evolver introduces a task-specific evaluation space for vulnerability reasoning.
This includes eight dimensions representing different categories of vulnerability-related context, including the candidate vulnerable API, usage snippets, exploit structure, debugger output, and code coverage. 
Table~\ref{tab:evolver_context_dimensions} describes these dimensions. 

For a failed exploit, the score for the prompt that produced it is given a score that estimates how useful the information corresponding to the dimension is for diagnosing the failure and revising the next prompt. 
This score does not measure whether the generated exploit satisfies the concrete triggering condition, which is not known at generation time. 
Instead, it estimates how likely the corresponding context dimension has been addressed by the exploit-generating prompt. 
\framework scores the usefulness of the available context along each dimension and generates a brief justification. 
The justifications are then used to produce natural language feedback for prompt evolution. 
They enforce a structure to the assessment of the context and help identify areas that need improvement. 

For configurations that use \pocgen as the Exploit Generator, a failed generation episode may produce multiple prompt candidates from different exploitation sources and refinement attempts.
We use these failed prompt candidates as the initial population for the Prompt Evolver and retain five semantically diverse candidates to cover representative failure modes while avoiding redundant LLM calls.
Each failed prompt candidate is represented together with the model response and the generated PoC exploit, which provides the input context for scoring, feedback generation, and prompt evolution.

Figure~\ref{fig:prompt-evolver} illustrates the two-stage workflow of the Prompt Evolver.
In the first stage, an LLM evaluates each failed prompt candidate using a rubric-based scoring prompt.
We design eight dimensions for assessing the usefulness of vulnerability-related context: 
candidate vulnerable API, 
vulnerability description, 
usage snippets, 
exploit skeleton, 
similar exploits, 
taint-path snippets, 
debugger output, and code-coverage signals.
For each criterion, the LLM assigns a score in the normalized range $[0,1]$ and returns a concise justification.
This constrained output format makes heterogeneous vulnerability-related context comparable across candidates.

\begin{table*}[t]
\centering
\small
\caption{Vulnerability-context dimensions for evaluating failed exploit-generation attempts.}
\label{tab:evolver_context_dimensions}
\begin{tabular}{lp{0.75\linewidth}}
\toprule
\textbf{Context dimension} & \textbf{Vulnerability-Related Context Assessed} \\
\midrule
Candidate Vulnerable API &
Information identifying a plausible API, method, or code location through which the vulnerable behavior may be reached.
The scoring model evaluates whether this information is sufficiently supported and useful for constructing an exploit that reaches the relevant execution path~\cite{you2017semfuzz,chen2024exploiting,kang2022test}. \\
\midrule
Vulnerability Description & Information describing the vulnerability type, trigger condition, and attacker-controlled input.
The scoring model evaluates whether this description is consistent with the available evidence and provides useful guidance for exposing the vulnerable behavior~\cite{you2017semfuzz,nong2025appatch}. \\
\midrule
Usage Snippets & Information showing how the candidate vulnerable API is installed, configured, and invoked in tests, documentation, or examples.
The scoring model evaluates whether these snippets provide useful call patterns and constraints for constructing an executable PoC exploit~\cite{kang2022test,chen2024exploiting,xie2022docter,wang2023carpetfuzz}.\\
\midrule
Exploit Skeleton & Information describing the high-level structure of a PoC exploit, including its setup, payload placement, vulnerable invocation, and validation.
The scoring model evaluates whether this structure provides useful guidance for producing a complete and executable exploit~\cite{marques2025automated,sun2025learning}.
Examples of the exploit skeletons used in our study are provided in the replication package\footnote{https://github.com/manh-td/pocevolve/blob/main/Appendix.pdf}. \\
\midrule
Similar Exploits & Information from prior exploits, including attack patterns, payload formats, call sequences, and validation strategies.
The scoring model evaluates whether these examples are relevant to the current vulnerability and provide useful patterns that can be adapted to it~\cite{10.5555/3495724.3495883,chen2024exploiting,deng2025chainfuzz}. \\
\midrule
Taint-path Snippets & Information showing how attacker-controlled data propagates from an input source to vulnerability-relevant operations.
The scoring model evaluates whether these data-flow paths provide useful guidance about where the payload should enter the program and how it should reach the vulnerable code~\cite{alhuzali2018navex,cassel2025nodemedic,marques2025automated}. \\
\midrule
Debugger Output & Information about runtime values, exceptions, crashes, and other observations produced by a failed exploit attempt.
The scoring model evaluates whether this output provides actionable evidence for revising the payload, API invocation, or execution setup~\cite{cassel2025nodemedic,deng2025chainfuzz,wang2018revery}. \\
\midrule
Code Coverage & Information about whether the generated exploit reaches vulnerability-relevant code and where its execution diverges.
\pocgen gets code coverage by driving Node's V8 Inspector Profiler directly around each PoC's execution in its validator.
The scoring model evaluates whether these signals provide actionable guidance for revising the entry point, payload, or trigger condition~\cite{yang20231dfuzz,cassel2025nodemedic,deng2025chainfuzz,wang2018revery,houis2026bullseye}. \\
\bottomrule
\end{tabular}
\end{table*}

The scoring process turns heterogeneous vulnerability-related context into structured optimization signals.
Rather than treating all available information as equally useful, the Prompt Evolver estimates how useful the information associated with each context dimension is for revising the failed generation attempt, 
and which are associated with irrelevant, redundant, or misleading context. 

In the second stage, the Prompt Evolver uses these scores to generate structured feedback and optimize prompts across iterations. Algorithm~\ref{alg:prompt-evolver} summarizes the overall evolution process. As depicted in the lower part of Figure~\ref{fig:prompt-evolver}, each prompt is paired with its generated exploit, Verifier result, usefulness scores, and feedback to form a candidate prompt state. The Prompt Evolver then applies Pareto candidate selection, detailed in Algorithm~\ref{alg:pareto-prompt-selection}, to retain only non-dominated prompts that achieve strong trade-offs across the vulnerability-context dimensions. This design prevents the evolution process from overfitting to a single feedback signal and preserves diverse prompt candidates that may support different exploit-generation strategies.

For feedback generation, the Prompt Evolver constructs a patch- and execution-aware diagnostic prompt. 
This prompt provides the failed prompt, model response, generated PoC exploit, vulnerability-fixing commit, vulnerability type, and the criterion usefulness scores. 
The LLM is instructed to 
identify why the attempt failed, and produce concise, actionable edits that should be made to the prompt. 
We constrain the feedback format to two parts: why the PoC exploit failed and how the user prompt should be changed, using operations such as adding, removing, or replacing prompt content.

Building on the selected Pareto-front candidates, an LLM performs reflective prompt evolution using a synthesis prompt. The prompt provides the current seed prompt, when available, together with Pareto-front prompt--feedback examples. The LLM is instructed to extract recurring failure patterns, such as missing code blocks, unclear payloads, incorrect API usage, excessive verbosity, or missing vulnerability-specific details, and synthesize one improved user prompt that directly addresses these failures while preserving the exploit-generation objective. The resulting prompt is then used in the next generation attempt, forming a closed-loop process that iteratively revises the prompt based on the assessed usefulness of vulnerability-related context.
The process terminates as soon as the Verifier confirms that a generated PoC successfully triggers the vulnerability or when the predefined evolution budget, set to 5 iterations in our experiments, is exhausted.

The Prompt Evolver complements runtime-based refinement with vulnerability-context-guided scoring and feedback.
While \textsc{GEPA} provides the general evolutionary optimization mechanism, our task-specific criteria define how vulnerability-related context is assessed and transformed into optimization signals.
The Prompt Evolver combines these signals with Verifier's feedback to evolve prompts toward executable exploit generation, particularly when execution feedback alone is insufficient to explain why an attempt fails.

\begin{algorithm}[t]
\caption{Prompt Evolver}
\label{alg:prompt-evolver}
\small
\begin{algorithmic}[1]
\Require Vulnerability context $\mathcal{C}$, diverse seed prompt groups ${\mathcal{P}_1,\ldots,\mathcal{P}_m}$, Exploit Generator $\mathcal{G}$, Verifier $\mathcal{V}$, vulnerability-context dimensions $\mathcal{K}$, iteration budget $T$
\Ensure Successful PoC exploit $e^*$ or best evolved prompt $p^*$

\State $\mathcal{A}_{\mathrm{all}} \gets \emptyset$

\ForAll{seed group $\mathcal{P}_i$}
\State $\mathcal{A} \gets \emptyset$

\ForAll{$p \in \mathcal{P}_i$} \Comment{evaluate pre-generated responses}
    \State $e \gets \textsc{Extract}(p.\mathit{response})$
    \State $r \gets \mathcal{V}(e)$
    \If{$r.\textsc{success}$}
        \State \Return $e$
    \EndIf
    \State $s \gets \textsc{Score}(p, e, \mathcal{C}, \mathcal{K})$
    \State $f \gets \textsc{Feedback}(p, e, r, \mathcal{C}, s)$
    \State $\mathcal{A} \gets \mathcal{A} \cup \{(p,e,r,s,f)\}$
    \State $\mathcal{A}_{\mathrm{all}} \gets \mathcal{A}_{\mathrm{all}} \cup \{(p,e,r,s,f)\}$
\EndFor

\For{$t = 1$ \textbf{to} $T$}
    \State $\mathcal{Q} \gets \textsc{ParetoCandidateSelection}(\mathcal{A}, \mathcal{K})$
    \State $p' \gets \textsc{EvolvePrompt}(\mathcal{C}, \mathcal{Q})$
    \State $e' \gets \mathcal{G}(p')$
    \State $r' \gets \mathcal{V}(e')$
    \If{$r'.\textsc{success}$}
        \State \Return $e'$
    \EndIf
    \State $s' \gets \textsc{Score}(p', e', \mathcal{C}, \mathcal{K})$
    \State $f' \gets \textsc{Feedback}(p', e', r', \mathcal{C}, s')$
    \State $\mathcal{A} \gets \mathcal{A} \cup \{(p',e',r',s',f')\}$
    \State $\mathcal{A}_{\mathrm{all}} \gets \mathcal{A}_{\mathrm{all}} \cup \{(p',e',r',s',f')\}$
\EndFor

\EndFor

\State $a^* \gets \arg\max_{a\in\mathcal{A}_{\mathrm{all}}}\textsc{AggregateScore}(a.s,a.r)$
\State \Return $a^*.p$
\end{algorithmic}
\end{algorithm}

\begin{algorithm}[t]
\caption{Pareto Candidate Selection}
\label{alg:pareto-prompt-selection}
\small
\begin{algorithmic}[1]
\Require Candidate states $\mathcal{A}=\{(p,e,r,s,f)\}$, vulnerability-context criteria $\mathcal{K}$
\Ensure Pareto-front candidate set $\mathcal{Q}$

\If{$\mathcal{A}=\emptyset$}
\State \Return $\emptyset$
\EndIf

\State $\mathcal{Q} \gets \emptyset$

\ForAll{$a_i \in \mathcal{A}$}
\State $\mathit{dominated} \gets \textsc{False}$
\ForAll{$a_j \in \mathcal{A}$}
\If{$a_i \neq a_j$ and
$\forall k\in\mathcal{K}: a_j.s[d]\geq a_i.s[d]$ and
$\exists k\in\mathcal{K}: a_j.s[d] > a_i.s[d]$}
\State $\mathit{dominated} \gets \textsc{True}$
\State \textbf{break}
\EndIf
\EndFor
\If{$\neg \mathit{dominated}$}
\State $\mathcal{Q} \gets \mathcal{Q} \cup \{a_i\}$
\EndIf
\EndFor

\State \Return $\mathcal{Q}$
\end{algorithmic}
\end{algorithm}

\section{Evaluation Settings}
\label{sec:settings}

\subsection{Research Questions}
To evaluate the effectiveness and practicality of \framework, we investigate the following research questions:

\begin{itemize}
\item \textbf{RQ1}: To what extent can \framework generate PoC exploits directly from vulnerability-fixing commits?
\item \textbf{RQ2}: How well does \framework perform when vulnerability report information is available?
\item \textbf{RQ3}: What trade-offs does \framework achieve between PoC exploit generation effectiveness, monetary cost, and runtime overhead?
\end{itemize}

RQ1 evaluates \framework in our proposed setting, where vulnerability reports are not provided. 
We run our experiments on SecBench.VFC.js, as described in Section~\ref{benchmark}. 
We compare \framework against \pocgen, which receives vulnerability information recovered from the fixing commit instead of a vulnerability report.
End-to-end success is measured using \textit{success rate}, defined as the proportion of vulnerabilities for which the generator produces an exploit validated by the verifier provided by Simsek et al.~\cite{simsek2025pocgen}.

RQ2 evaluates \framework against \pocgen on SecBench.js when vulnerability report information is available. 
In this setting, \framework uses report-provided vulnerability information directly and skips the Commit Analyzer.
This setting examines whether prompt evolution can further improve PoC exploit generation when richer vulnerability context is provided. 

RQ3 examines the practicality of \framework. While prompt evolution may improve exploit-generation effectiveness, it also introduces additional LLM calls, which can increase monetary cost and runtime. We therefore measure the trade-offs between success rate, cost, and runtime overhead to determine whether the effectiveness gains remain practical.

\subsection{Experimental Setup}

\subsubsection{Benchmark}
\label{benchmark}

We construct our benchmark based on SecBench.js~\cite{bhuiyan2023secbench}.
To ensure a fair comparison with \pocgen, we use the same subset of vulnerabilities from SecBench.js that were evaluated by Simsek et al.~\cite{simsek2025pocgen}.
We refer to this subset of vulnerabilities as \textbf{SecBench.js}.
Table~\ref{tab:vfcs_distribution_combined} summarizes the distribution of vulnerability types in SecBench.js.

To evaluate exploit generation directly from code changes, we further filter SecBench.js to retain only vulnerabilities with corresponding vulnerability-fixing commits or pull requests provided by Bhuiyan et al.~\cite{bhuiyan2023secbench}.
Among these, 219 vulnerabilities are associated with vulnerability-fixing commits.
We extract the commit messages and code changes from these commits, 
and exclude any vulnerabilities whose associated commit contains no source code changes, such as those involving only documentation or test modifications.
We further eliminate duplicate vulnerabilities. 
Although such duplicates may correspond to multiple vulnerability types, they typically share identical functionality and exploit patterns. 
After deduplication, the final benchmark consists of 190 vulnerabilities with source code changes, which we refer to as \textbf{SecBench.VFC.js}. 
SecBench.VFC.js is used in RQ1, while SecBench.js is used in RQ2.
Table~\ref{tab:vfcs_distribution_combined} summarizes the distribution of vulnerability types before and after filtering.

\begin{table}[t]
\centering
\caption{Distribution of vulnerability types in SecBench.js and SecBench.VFC.js.}
\label{tab:vfcs_distribution_combined}
\scalebox{0.9}{
\begin{tabular}{lccc ccc}
\toprule
\multirow{2}{*}{\textbf{Vulnerability Type}} 
& \multicolumn{3}{c}{\textbf{SecBench.js}}
& \multicolumn{3}{c}{\textbf{SecBench.VFC.js}} \\
\cmidrule(lr){2-4} \cmidrule(lr){5-7}
& \textbf{GHSA} & \textbf{Snyk} & \textbf{Total}
& \textbf{GHSA} & \textbf{Snyk} & \textbf{Total} \\
\midrule
Path Traversal        & 85  & 82 & 167 & 8   & 2  & 10 \\
Prototype Pollution   & 136 & 44 & 180 & 61  & 20 & 81 \\
Command Injection     & 66  & 24 & 90  & 24  & 4  & 28 \\
Code Injection        & 26  & 10 & 36  & 9   & 4  & 13 \\
ReDoS                 & 59  & 27 & 86  & 40  & 18 & 58 \\
\midrule
Total                 & 372 & 187 & 559 & 142 & 48 & 190 \\
\bottomrule
\end{tabular}
}
\end{table}

\subsubsection{Large Language Models}

We evaluate \textit{GPT-4o-mini} and \textit{Qwen3.7-Plus} as the underlying LLMs.
GPT-4o-mini is selected as it was used in the original \pocgen evaluation, enabling a direct comparison.
We additionally evaluate both \framework configurations with Qwen3.7-Plus in RQ1 and RQ3 to examine how effectiveness, monetary cost, and runtime vary across underlying LLMs and Exploit Generators.
Within each configuration, the same underlying LLM is used for the Commit Analyzer, Exploit Generator, and Prompt Evolver.
We use a temperature of 1, consistent with the original \pocgen configuration.
Although this setting permits stochastic outputs, three repeated runs showed little variation in the overall results.
GPT-4o-mini is accessed through the OpenAI API, while Qwen3.7-Plus is accessed through the OpenRouter API.

\subsubsection{Baselines}
\textbf{Large Language Model.} Given that LLMs have demonstrated strong capabilities across a wide range of software engineering tasks~\cite{ouedraogo2026prompt, zhou2024large, pearce2023examining},
we use direct LLM prompting as a baseline to evaluate whether an LLM can generate PoC exploits from vulnerability-fixing commits without the specialized components of \framework.
The LLM receives the vulnerability-fixing commit and accessible repository context and is prompted directly to generate a PoC exploit.
This baseline does not use \pocgen's program-analysis-assisted generation process or \framework's Prompt Evolver.

\textbf{\pocgen.} 
\pocgen has been shown to outperform prior npm vulnerability exploitation tools, including NodeMedic-FINE~\cite{cassel2025nodemedic}, 
FAST~\cite{kang2023scaling},
and Explode.js~\cite{marques2025automated}. 
We use \pocgen as the state-of-the-art baseline exploit generator. 

For RQ1, we adapt \pocgen to the simulated use case of generating exploits before a complete vulnerability report is available.  
While \pocgen requires detailed vulnerability information, such information is unavailable in this setting.   
Instead of using the vulnerability reports, we provide the commit-recovered vulnerability information inferred by the Commit Analyzer.
For RQ2, we use the original implementation of \pocgen as the vulnerability reports are provided. 
However, all approaches are evaluated using our corrected verifier, which addresses the false-positive behavior identified for Command Injection vulnerabilities.

\subsubsection{Testbed and Verifier}
\label{sec:Verifier}

To ensure a fair comparison with prior work, we use the execution environment provided by \pocgen's replication package.
We reuse their scripts to download vulnerable packages and construct Docker containers for evaluation. 
This allows an evaluation of \framework under the same conditions.
For exploit validation, we use the verifier functions provided by \pocgen, with a correction to the Command Injection verifier that prevents false-positive exploits from being accepted.
The same corrected verifier is applied to all evaluated approaches.
They automatically determine whether a generated exploit successfully triggers the target vulnerability, ensuring consistent evaluation across all approaches.

\subsubsection{Implementation and Hardware}
\label{sec:implementation}

Our approach is implemented in Python~3. We use the OpenAI API to interact with GPT-family models and the GitHub REST API\footnote{\url{https://docs.github.com/en/rest}} to retrieve repository data. We build on the replication package of \pocgen, extending and modifying it to support our experimental settings while preserving its default hyperparameters, including up to 30 iterative refinement rounds and a temperature of 1, which \framework also uses across all OpenAI API interactions.
Since we adopt \pocgen's Verifier and generation interface, generated exploits follow the same output convention and are wrapped inside a single entry function \texttt{exploit()} for compatibility with the evaluation framework.
We provide all prompts used in our implementation in the replication package\footnote{https://github.com/manh-td/pocevolve/blob/main/Appendix.pdf}.

During prompt evolution, we initialized the Prompt Evolver with five semantically diverse prompts selected from failed \pocgen generations and performed five evolution iterations. For \framework (LLM), we limited the generator's refinement loop
to five iterations to balance between success rate and cost.

All experiments are conducted on a server equipped with AMD EPYC 7763 64-Core Processor @ 3.53 GHz, 251\,GB RAM, and running Ubuntu 22.04.5 LTS. To improve efficiency, we execute multiple experiments in parallel. To avoid race conditions during file operations, we run 10 independent pipelines, each within a separate Docker container.

\section{Experimental Results}
\label{sec:results}

\subsection{RQ1: To what extent can \framework generate PoC exploits directly from vulnerability-fixing commits?}

We evaluate \framework and the baselines on SecBench.VFC.js in our proposed setting using two underlying LLMs: GPT-4o-mini and Qwen3.7-Plus.
For both LLMs, we evaluate four configurations: LLM, \framework (LLM), \pocgen, and \framework.
Within each comparison, \framework and its corresponding baseline use the same underlying LLM, allowing us to isolate the contribution of prompt evolution.

Table~\ref{tab:rq1_results} presents the overall success rates.
With GPT-4o-mini, the LLM baseline achieves a success rate of 19.5\%.
Applying \framework's Prompt Evolver to the LLM generator increases the success rate to 28.4\%, representing a relative improvement of 45.9\%.
\pocgen achieves a success rate of 48.4\%.
By applying prompt evolution to \pocgen, \framework further increases the success rate to 58.4\%.
This corresponds to absolute improvements of 10.0 percentage points over \pocgen and 38.9 percentage points over the LLM baseline, equivalent to relative improvements of 20.7\% and 200.0\%\footnote{Relative improvements are calculated using the underlying success counts as $(\text{new}-\text{old})/\text{old}\times100\%$; for example, $(111-37)/37\times100\%=200.0\%$.}, respectively.

With Qwen3.7-Plus, the LLM baseline achieves a success rate of 77.9\%.
Applying the Prompt Evolver to the direct LLM generator increases the success rate to 85.3\%, corresponding to an absolute improvement of 7.4 percentage points and a relative improvement of 9.5\%.
For the \pocgen-based configurations, \pocgen achieves a success rate of 64.2\%, while the full \framework achieves 79.5\%. 
This corresponds to an absolute improvement of 15.3 percentage points and a relative improvement of 23.8\%.
Thus, prompt evolution improves both evaluated Exploit Generators with Qwen3.7-Plus.

\begin{figure*}[t]
    \includegraphics[width=0.49\linewidth]{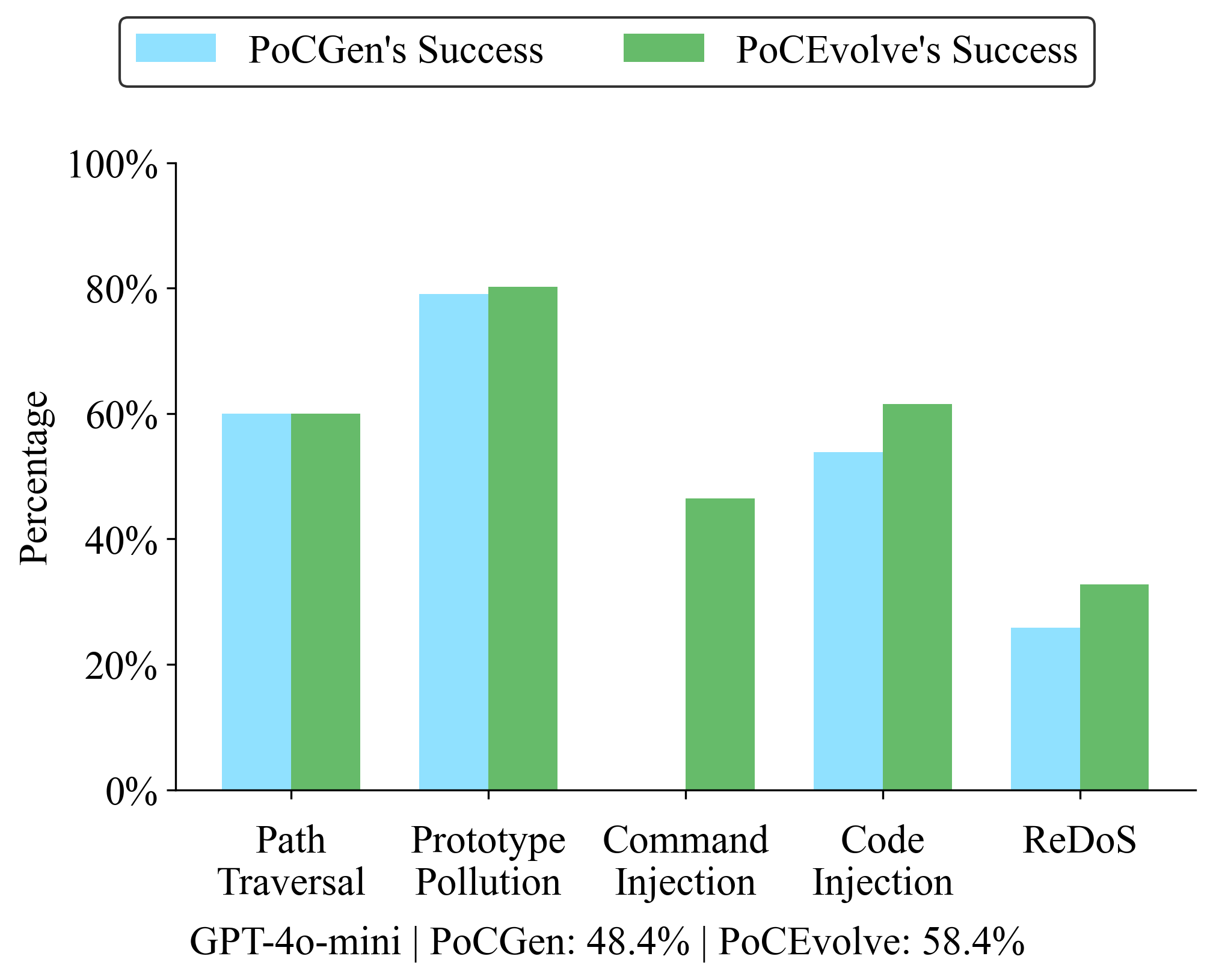}
    \includegraphics[width=0.49\linewidth]{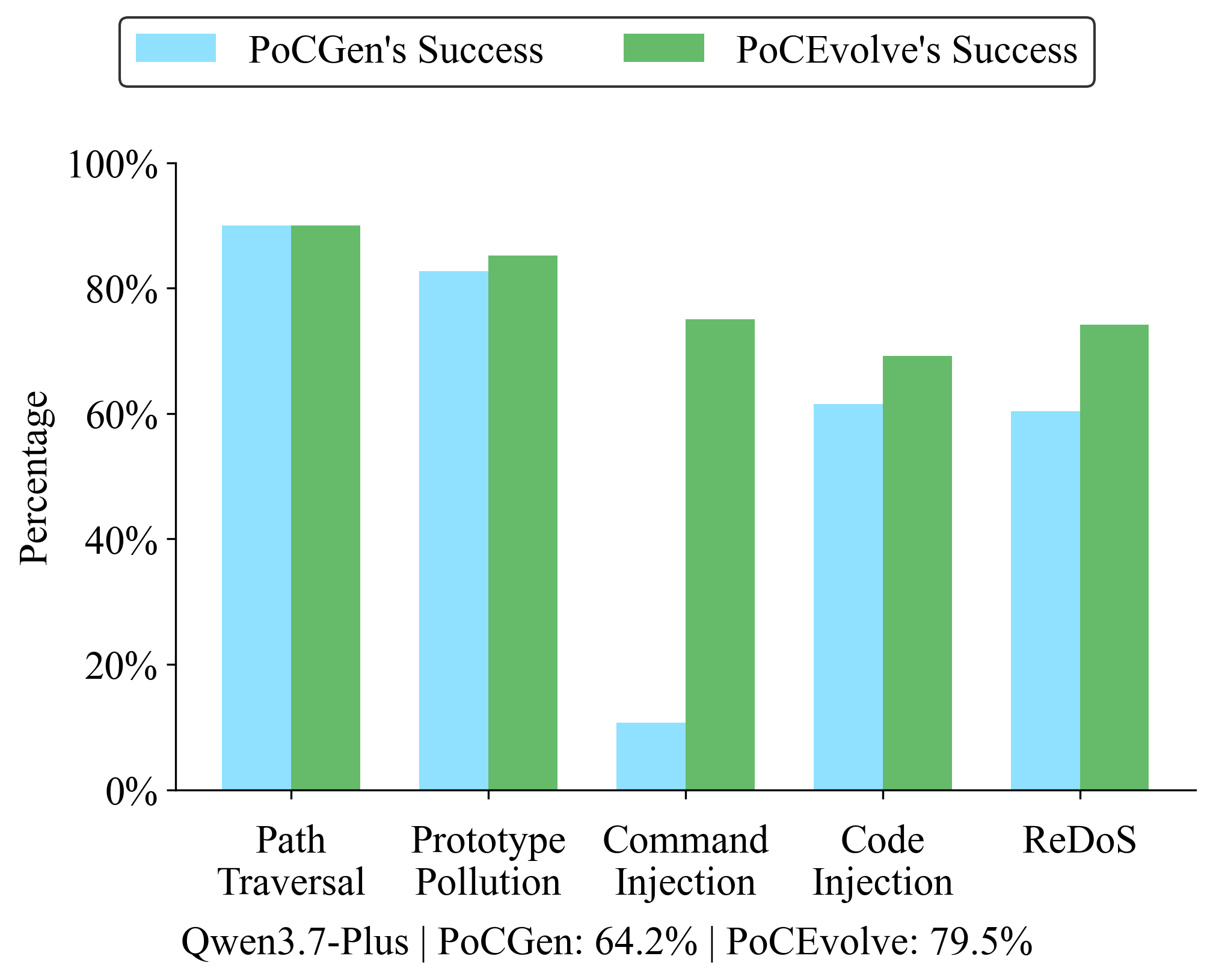}
    \caption{PoC exploit generation success rates by vulnerability type using GPT-4o-mini (left panel) and Qwen3.7-Plus (right panel) in our proposed setting. Within each panel, the left bar represents \pocgen, while the right bar represents \framework.}
    \label{fig:approach_results}
\end{figure*}

Figure~\ref{fig:approach_results} reports the results by vulnerability type. 
\framework outperforms \pocgen for four of the five vulnerability types under both LLMs and achieves the same success rate for Path Traversal.
The largest improvement is observed for Command Injection.
With GPT-4o-mini, \pocgen fails to generate any successful Command Injection exploits, whereas \framework succeeds on nearly half of these vulnerabilities.
With Qwen3.7-Plus, \framework increases the Command Injection success rate from approximately 11\% to 75\%.
Command Injection often requires simultaneously identifying the correct vulnerable API, constructing an appropriate payload, and satisfying environment-specific triggering conditions.
Prompt evolution uses feedback derived from failed attempts to revise the API, payload, and environment information expressed in subsequent prompts.

These results show that vulnerability-fixing commits and repository context can provide sufficient information to generate executable PoC exploits.
With GPT-4o-mini, \framework (LLM) and the full \framework achieve success rates of 28.4\% and 58.4\%, respectively.
With Qwen3.7-Plus, these configurations achieve success rates of 85.3\% and 79.5\%, respectively.
Prompt evolution improves its corresponding Exploit Generator under both underlying LLMs.

These findings suggest that vulnerability-fixing commits can support the reconstruction of concrete vulnerability-triggering behavior and motivate evaluating future exploit generators under a patch-only setting rather than assuming that a detailed report is already available.

\begin{center}
\begin{tcolorbox}[
  colback=gray!10,
  colframe=darkgray,
  boxrule=1.5pt,
  arc=4pt,
  width=0.98\linewidth
]
\textbf{Answer to RQ1.}
\framework successfully generates PoC exploits from vulnerability-fixing commits and improves both evaluated Exploit Generators.
With GPT-4o-mini, \framework (LLM) improves over direct LLM prompting by 45.9\%, while the full \framework improves over \pocgen by 20.7\%.
With Qwen3.7-Plus, \framework (LLM) improves over direct LLM prompting by 9.5\%, while the full \framework improves over \pocgen by 23.8\%.
These findings establish patch-only exploit generation as a feasible and practically important evaluation setting and demonstrate that the benefits of prompt evolution generalize across different underlying LLMs and Exploit Generators.
\end{tcolorbox}
\end{center}

\begin{table*}[t]
\centering
\caption{PoC exploit generation results on SecBench.VFC.js. \framework (LLM) denotes a variant of \framework that does not use \pocgen as its exploit-generation component, instead relying on direct LLM prompting.}
\label{tab:rq1_results}
\small
\begin{tabular}{llcc}
\toprule
\textbf{Underlying LLM} &
\textbf{Method} &
\textbf{Success} &
\textbf{Improvement} \\
\midrule

\multirow{4}{*}{GPT-4o-mini}
& LLM
& 19.5\%
& -- \\

& \framework (LLM)
& 28.4\%
& +45.9\% over LLM \\

& \pocgen
& 48.4\%
& -- \\

& \textbf{\framework}
& \textbf{58.4\%}
& \begin{tabular}[c]{@{}c@{}}
  \textbf{+20.7\% over \pocgen} \\
  \textbf{+200.0\% over LLM}
  \end{tabular} \\

\midrule

\multirow{4}{*}{Qwen3.7-Plus}
& LLM
& 77.9\%
& -- \\

& \textbf{\framework (LLM)}
& \textbf{85.3\%}
& \textbf{+9.5\% over LLM} \\

& \pocgen
& 64.2\%
& -- \\

& \framework
& 79.5\%
& +23.8\% over \pocgen \\

\bottomrule
\end{tabular}
\end{table*}

\subsection{RQ2: How well does \framework perform when vulnerability report information is available?}

We evaluate \framework and \pocgen on SecBench.js with vulnerability report information provided to both approaches.
For this RQ, we use only GPT-4o-mini as the underlying LLM because it is the model used in the original \pocgen evaluation.
This choice enables a direct comparison with \pocgen while keeping the underlying model and experimental configuration consistent.
Therefore, the observed performance differences primarily reflect the contribution of \framework's prompt evolution rather than differences between LLMs.

\begin{figure}[t]
    \centering
    \includegraphics[width=\linewidth]{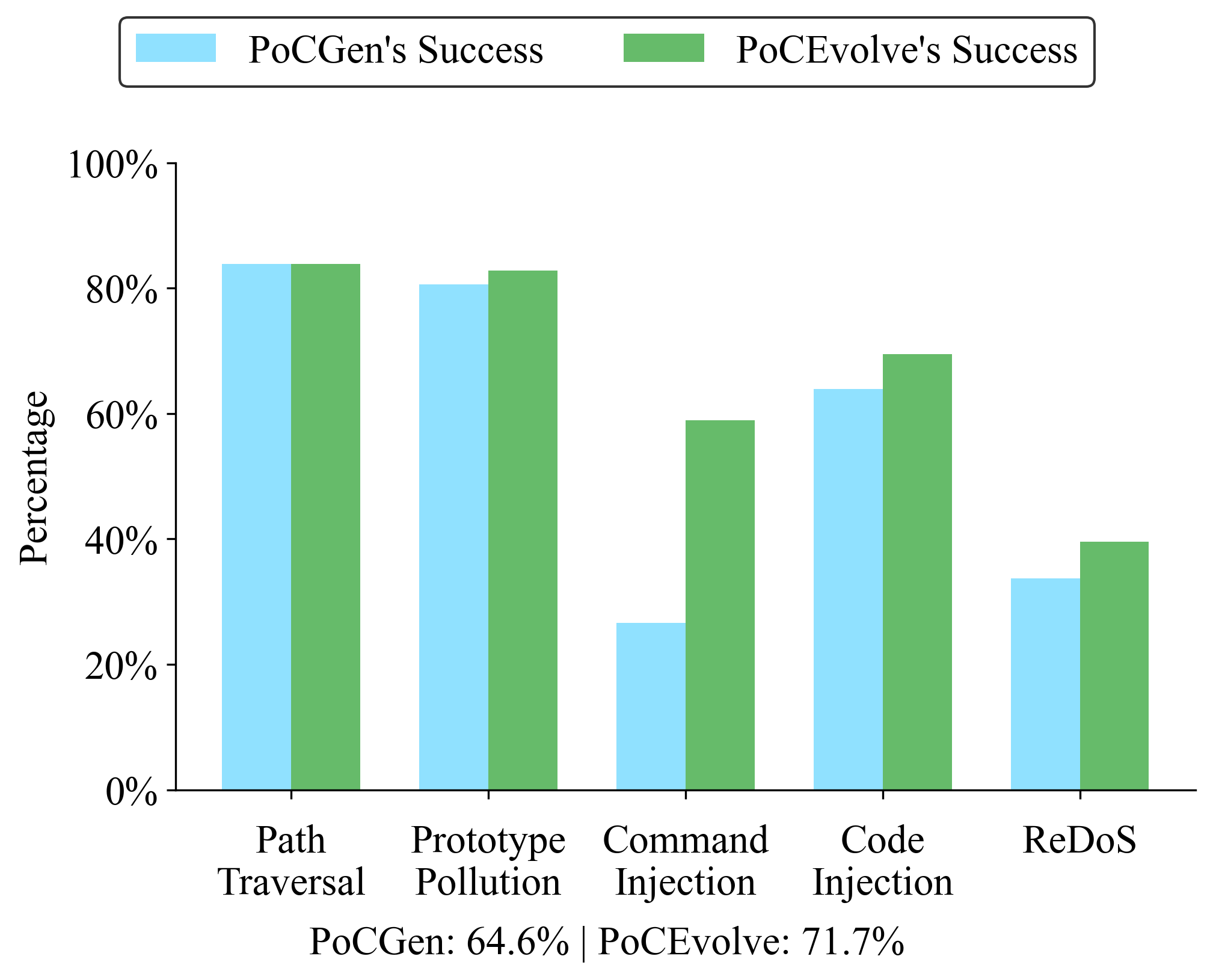}
    \caption{Experimental results by vulnerability type in the Report-guided setting, i.e., detailed vulnerability reports are available. For each vulnerability type, the left bar shows the success rate of \pocgen, while the right bar shows the success rate of \framework.}
    \label{fig:rq2_oneday}
\end{figure}

Figure~\ref{fig:rq2_oneday} presents the results in the report-guided setting, where vulnerability reports are available. 
It shows that  
\framework achieves a success rate of 71.7\%,
while 
\pocgen achieves a PoC exploit generation success rate of 64.6\%.
This improvement is particularly pronounced for Command Injection vulnerabilities. This result suggests that prompt evolution can recover unsuccessful generations for which \pocgen's predefined runtime feedback is insufficient. 

Detailed vulnerability reports do not eliminate failures in exploit generation. Even in the report-guided setting, \framework improves the success rate from 64.6\% to 71.7\%, suggesting that some unsuccessful \pocgen executions contain information that can be recovered through subsequent feedback-guided prompt revision. 
Thus, runtime feedback expressed through fixed, predefined hints does not fully exploit the information available in failed attempts. Structuring the assessment of vulnerability-related context and generating attempt-specific feedback can improve refinement even when the initial vulnerability description is already detailed.

Overall, these results show that prompt evolution is able to recover helpful information even if detailed vulnerability reports are available. 
Although the report-guided setting provides rich vulnerability information, 
\framework can still generate useful attempt-specific feedback from unsuccessful generations and use it to guide subsequent refinement.
This guides \pocgen toward valid PoC exploits.

\begin{center}
\begin{tcolorbox}[
  colback=gray!10,
  colframe=darkgray,
  boxrule=1.5pt,
  arc=4pt,
  width=0.98\linewidth
]
\textbf{Answer to RQ2.}
When vulnerability reports are available, 
\framework increases the PoC exploit generation success rate from 64.6\% to 71.7\%, which is an improvement of 11.1\% over \pocgen.
These results demonstrate that prompt evolution improves PoC exploit generation beyond the {proposed} setting.
\end{tcolorbox}
\end{center}

\subsection{RQ3: What trade-offs does \framework achieve between PoC exploit generation effectiveness, monetary cost, and runtime overhead?}

\begin{table*}[t]
\centering
\caption{Monetary cost and runtime per vulnerability across the evaluated configurations and settings. \framework (LLM) denotes a variant of \framework that does not use \pocgen as its exploit-generation component, instead relying on direct LLM prompting.}
\label{tab:rq3_cost_runtime_combined}

\resizebox{\textwidth}{!}{%
\begin{tabular}{l l l l c l c c c c c c c c}
\toprule
\multirow{2}{*}{\textbf{Approach}} &
\multirow{2}{*}{\textbf{Model}} &
\multirow{2}{*}{\textbf{Setting}} &
\multirow{2}{*}{\textbf{Dataset}} &
\multirow{2}{*}{\textbf{Size}} &
\multirow{2}{*}{\textbf{Subset}} &
\multicolumn{4}{c}{\textbf{Mean cost per vulnerability (USD)}} &
\multicolumn{4}{c}{\textbf{Mean runtime per vulnerability (H:MM:SS)}} \\
\cmidrule(lr){7-10} \cmidrule(lr){11-14}
& & & & & &
\textbf{Analyzer} &
\textbf{Generator} &
\textbf{Evolver} &
\textbf{Total} &
\textbf{Analyzer} &
\textbf{Generator} &
\textbf{Evolver} &
\textbf{Total} \\
\midrule

\framework &
GPT-4o-mini &
Report-guided &
SecBench.js &
559 &
Successful only &
-- & 0.0077 & 0.0237 & 0.0313 &
-- & 0:04:38 & 0:04:46 & 0:09:24 \\

& & & & &
All &
-- & 0.0300 & 0.0530 & 0.0830 &
-- & 0:13:12 & 0:08:32 & 0:21:43 \\

\midrule

\framework (LLM) &
GPT-4o-mini &
Our proposed setting &
SecBench.VFC.js &
190 &
Successful only &
0.0003 & 0.0007 & 0.0152 & 0.0162 &
0:00:02 & 0:00:19 & 0:03:10 & 0:03:29 \\

& & & & &
All &
0.0003 & 0.0029 & 0.0260 & 0.0292 &
0:00:02 & 0:01:05 & 0:04:46 & 0:05:52 \\

\midrule

\framework &
GPT-4o-mini &
Our proposed setting &
SecBench.VFC.js &
190 &
Successful only &
0.0003 & 0.0077 & 0.0174 & 0.0253 &
0:00:02 & 0:05:26 & 0:04:31 & 0:09:57 \\

& & & & &
All &
0.0003 & 0.0363 & 0.0529 & 0.0895 &
0:00:02 & 0:15:06 & 0:11:14 & 0:26:19 \\

\midrule

\framework (LLM) &
Qwen3.7-Plus &
Our proposed setting &
SecBench.VFC.js &
190 &
Successful only &
0.0018 & 0.0068 & 0.0631 & 0.0717 &
0:02:00 & 0:02:17 & 0:25:07 & 0:29:24 \\

& & & & &
All &
0.0020 & 0.0113 & 0.1081 & 0.1214 &
0:02:00 & 0:04:32 & 0:38:08 & 0:44:40 \\

\midrule

\framework &
Qwen3.7-Plus &
Our proposed setting &
SecBench.VFC.js &
190 &
Successful only &
0.0018 & 0.0350 & 0.1057 & 0.1425 &
0:02:00 & 0:12:07 & 0:17:15 & 0:31:22 \\

& & & & &
All &
0.0020 & 0.0690 & 0.1782 & 0.2492 &
0:02:00 & 0:19:49 & 0:31:06 & 0:52:54 \\

\bottomrule
\end{tabular}%
}
\end{table*}

Table~\ref{tab:rq3_cost_runtime_combined} presents the monetary cost and runtime per vulnerability across the evaluated configurations.
Regarding monetary cost, with GPT-4o-mini in the proposed setting, \framework (LLM) costs \$0.0162 per successful vulnerability and \$0.0292 across all evaluated vulnerabilities, while the full \framework costs \$0.0253 and \$0.0895, respectively.
The overall cost of the full GPT-4o-mini configuration is 2.47 times its generator-only cost of \$0.0363.
In the report-guided setting, the full \framework costs \$0.0313 per successful vulnerability and \$0.0830 across all evaluated vulnerabilities.

With Qwen3.7-Plus in the proposed setting, \framework (LLM) costs \$0.0717 per successful vulnerability and \$0.1214 across all evaluated vulnerabilities.
The full \framework costs \$0.1425 and \$0.2492, respectively.
Thus, compared with the full Qwen3.7-Plus configuration, \framework (LLM) reduces the total cost by 49.7\% for successful vulnerabilities and by 51.3\% across all evaluated vulnerabilities.

Regarding runtime, with GPT-4o-mini in the proposed setting, \framework (LLM) requires 3 minutes and 29 seconds per successful vulnerability and 5 minutes and 52 seconds across all evaluated vulnerabilities.
The full \framework requires 9 minutes and 57 seconds and 26 minutes and 19 seconds, respectively.
In the report-guided setting, the full \framework requires 9 minutes and 24 seconds per successful vulnerability and 21 minutes and 43 seconds across all evaluated vulnerabilities.

With Qwen3.7-Plus, \framework (LLM) requires 29 minutes and 24 seconds per successful vulnerability and 44 minutes and 40 seconds across all evaluated vulnerabilities.
The full \framework requires 31 minutes and 22 seconds and 52 minutes and 54 seconds, respectively.
Compared with the full Qwen3.7-Plus configuration, \framework (LLM) reduces runtime by 1 minute and 58 seconds for successful vulnerabilities and by 8 minutes and 14 seconds across all evaluated vulnerabilities.
The Prompt Evolver accounts for most of the total cost and runtime of \framework (LLM), for which the direct LLM generator itself is comparatively inexpensive.

The evaluated configurations exhibit different trade-offs between effectiveness, monetary cost, and runtime. With GPT-4o-mini, \framework (LLM) achieves a success rate of 28.4\% at an average cost of \$0.0292 and a runtime of 5 minutes and 52 seconds per evaluated vulnerability, while the full \framework achieves 58.4\% at \$0.0895 and 26 minutes and 19 seconds. 
With Qwen3.7-Plus, \framework (LLM) achieves 85.3\% at \$0.1214 and 44 minutes and 40 seconds, while the full \framework achieves 79.5\% at \$0.2492 and 52 minutes and 54 seconds.
These results show that both the underlying LLM and the choice of Exploit Generator materially affect the effectiveness and resource requirements of prompt evolution.
The additional cost and runtime introduced by prompt evolution should be considered relative to the security window in which the technique is intended to operate.

Our empirical study in Section~\ref{sec:empirical_study} indicates that the median interval between the first vulnerability-fixing commit and vulnerability disclosure is 18 days.
Across the proposed-setting configurations, the mean runtime ranges from 5 minutes and 52 seconds to 52 minutes and 54 seconds per evaluated vulnerability.
Although processing a large collection of vulnerabilities would require parallel execution, these per-vulnerability runtimes are substantially shorter than the median patch-to-disclosure interval.
This suggests that automated PoC synthesis can be performed within the patch-to-disclosure window, reinforcing the need for rapid patch deployment.

\begin{center}
\begin{tcolorbox}[
  colback=gray!10,
  colframe=darkgray,
  boxrule=1.5pt,
  arc=4pt,
  width=0.98\linewidth
]
\textbf{Answer to RQ3.}
Prompt evolution improves PoC exploit generation but introduces additional monetary and runtime overhead.
In our proposed setting, \framework (LLM) achieves success rates of 28.4\% and 85.3\% with GPT-4o-mini and Qwen3.7-Plus, at average costs of \$0.0292 and \$0.1214 and runtimes of 5 minutes and 52 seconds and 44 minutes and 40 seconds per evaluated vulnerability, respectively.
The full \framework achieves success rates of 58.4\% and 79.5\%, at average costs of \$0.0895 and \$0.2492 and runtimes of 26 minutes and 19 seconds and 52 minutes and 54 seconds, respectively.
When provided with detailed vulnerability reports, the full \framework improves over \pocgen by 11.1\%, at \$0.0830 and 21 minutes and 43 seconds per evaluated vulnerability.
\end{tcolorbox}
\end{center}

\section{Discussion}
\label{sec:discussion}

\subsection{Qualitative Analysis}
\label{subsec:qualitative-analysis}

\subsubsection{Success Case}
CVE-2019-10750 is a prototype pollution vulnerability in \texttt{deeply}, a library for recursively merging JavaScript objects.
The vulnerability-fixing commit introduces a check that rejects an enumerable \texttt{\_\_proto\_\_} key in an argument to the \texttt{reduceObject} function. 
If the \texttt{\_\_proto\_\_} key in an input object can be treated as enumerable, an attacker can pollute the base JavaScript object prototype to introduce unsafe behaviors. 

\framework initially generates unsuccessful exploits.
One such exploit used \texttt{\_proto\_} (with single underscores instead of double underscores) as the payload key. 
\framework's feedback identified the error: 
\textit{``Based on the structured payload using \texttt{\_proto\_} instead of the correct \texttt{\_\_proto\_\_}, hence the exploit did not trigger the expected prototype pollution. Add clarification to use \texttt{\_\_proto\_\_} as the key in the payload, ensuring the user understands the necessary structure for prototype pollution.''} and 
this is used to automatically improve the exploit generation prompt. 
Other exploit candidates, shown in  Listing~\ref{lst:deeply-fail}, fail to target the right vulnerable API, and instead targets the  \texttt{deeply.adapters.object} and \texttt{deeply.adapters.functionsExtend}, 
which do not trigger the vulnerable execution path. 
Several candidate exploits use payloads with the right \texttt{\_\_proto\_\_} key, but without preserving it as an enumerable property. 
\framework requires multiple iterations to accumulate enough information about each dimension required for the exploit. 
In particular, the feedback along some dimensions can be incorrect, such as those caused by the LLM's imperfect analysis, including cases when it incorrectly attributes an error to unrelated code. 
This requires multiple iterations for \framework to receive more feedback that accounts for these errors. 

As shown in Listing~\ref{lst:deeply-exploit}, using the feedback accumulated across multiple iterations, \framework changes its strategy. 
It invokes the public \texttt{deeply()} API with a payload created by \texttt{JSON.parse}.
The use of the public API routes it to the vulnerable code in \texttt{reduceObject}.
The use of \texttt{JSON.parse} allows it to preserve \texttt{\_\_proto\_\_} as an enumerable property. 
Such behavior demonstrates how multiple iterations are necessary to derive and refine the information for successful exploitation.
In particular, these pieces of information were not directly present in the vulnerability-fixing commit. 

\begin{listing}[!t]
\begin{minted}[frame=lines,framesep=2mm,fontsize=\scriptsize,breaklines=true]{javascript}
// Incorrect key
deeply.adapters.object({}, {_proto_: {exploited: true}}, (a, b) => b);
// Object-literal payload
deeply.adapters.object({}, {"__proto__": {exploited: true}}, (a, b) => b);
// Adapter incompatible with object inputs
deeply.adapters.functionsExtend({}, {"__proto__": {exploited: true}});
\end{minted}
\caption{Representative unsuccessful \framework exploits. }
\label{lst:deeply-fail}
\end{listing}

\begin{listing}[!t]
\begin{minted}[frame=lines,framesep=2mm,fontsize=\scriptsize,breaklines=true]{javascript}
const deeply = require('deeply');
const payload = JSON.parse('{"__proto__": {"exploited": true}}');
const target = {};
deeply(target, payload);
\end{minted}
\caption{Successful \framework exploit.}
\label{lst:deeply-exploit}
\end{listing}

\subsubsection{Analysis of Failure Cases}
We analyze the cases where \framework fails to generate an exploit, and identify two main failure modes.

\textbf{LLM context length limitations interrupt optimization.}
\framework can also fail when the context accumulated across iterations exceeds the LLM's maximum context length.
Since each iteration appends the prior prompt, exploit attempt, and execution feedback rather than replacing it, the size of the input grows monotonically as more iterations are attempted.
One common issue is that vulnerabilities requiring several refinement iterations, or involving verbose source code and execution logs, accumulate context that exceeds this limit before the optimization loop reaches its maximum number of iterations.
Once this limit is reached, the underlying LLM call fails outright, regardless of whether earlier iterations were converging toward a working exploit.
For example, in SNYK-JS-CODEMIRROR-1016937~\cite{snyk_codemirror_1016937}, repeated iterations accumulate prior attempts and execution logs until the prompt exceeds the token limit of the underlying LLM, causing the call to fail before optimization completes.
As a result, \framework fails to produce a working exploit for this vulnerability.

\textbf{Failures in the underlying exploit generator propagate to \framework.}
\framework can also be misled when the underlying Exploit Generator, e.g., \pocgen, incorrectly identifies vulnerability information.
If the initial vulnerability analysis is incorrect, \framework does not always recover from it.
One common issue is that the vulnerable API is inaccurately identified.
As the ground-truth information is not provided, \framework prompt evolver does not always succeed in identifying that this dimension of the exploit is incorrect as long as a plausible API is present.
For example, in CVE-2021-23436~\cite{snyk_immer_1540542}, the ground-truth vulnerability affects the \texttt{applyPatches} API, 
while \pocgen focuses on \texttt{produce}, Immer's standard API. 
As the use of a standard API looked like a plausible entry point, \framework fails to identify that this dimension of the exploit is not correct.

\subsection{Threats to Validity}
One threat to internal validity is implementation mistakes. 
To prevent errors, we have double-checked our implementation.
Another possible threat is that LLMs are recalling exploits from training data instead of synthesizing them. 
However, this threat is mitigated as we have used both a standard LLM approach and \textsc{PoCGen} as a baseline, which would obtain the same results as \framework if that were the case. 

Another threat to internal validity is the LLM context-length limit, as \framework accumulates substantial context across iterations, potentially interrupting its execution. However, this occurred no more than four times in any GPT-4o-mini setting and was not observed with Qwen3.7-Plus, which has a larger context window.

One threat to external validity is our focus on the npm ecosystem.
For a fair comparison against \pocgen, our experiments used the same experimental setup as Simsek et al.~\cite{simsek2025pocgen}.
This may prevent our findings from generalizing to other languages.
However, this is mitigated as the design of our approach is language-agnostic and our experiments include vulnerabilities of a range of different projects.  

One threat to construct validity is the choice of oracles for detecting the success of an exploit. 
This is mitigated as we adopted the same oracle-based evaluation protocol as Simsek et al.~\cite{simsek2025pocgen} which ensures a fair comparison. 
We also manually validated the verifiers and benchmarking code, including reproducing the results of Simsek et al.~\cite{simsek2025pocgen}. 
Our manual analysis was thorough, including identifying a flaw in the verifier from \textsc{PoCGen} for Command Injection vulnerabilities.\footnote{We have reported this issue to the authors of PoCGen.}  
As such, there is minimal threat to construct validity.

\section{Related Work}
\label{sec:related-work}

\textbf{Just-in-Time Vulnerability and Defect Prediction.}
Just-in-time (JIT) defect prediction classifies whether a commit introduces a bug using commit-level features, code diffs, and increasingly large language models~\cite{ni2022just, chen2024jit, zhou2025bridging, kamei2012large, fukushima2014empirical, zeng2021deep, hoang2020cc2vec, hoang2019deepjit}.
JIT vulnerability prediction extends this to security-critical changes, flagging commits likely to introduce exploitable weaknesses~\cite{nguyen2025toward, perl2015vccfinder, pan2023fine, li2023commit, minh2021deepcva}.
Regression bug detection similarly targets change-inducing defects at the commit level~\cite{pradel2025testora}. 
These approaches produce a risk score or label, but do not offer executable evidence that a vulnerability is exploitable. 
\framework automatically generates PoC exploits that confirm a vulnerability is real and reachable.

\textbf{Detecting and Exploiting Node.js Vulnerabilities.}
Prior tools for npm vulnerability detection and exploitation include NODEMEDIC-FINE~\cite{cassel2025nodemedic}, which uses dynamic taint analysis with fuzzer-generated inputs; FAST~\cite{kang2023scaling}, which traverses enriched control flow graphs to collect sink-reaching constraints; and \textsc{Explode.js}~\cite{marques2025automated}, which synthesizes guaranteed side-effect-producing exploits for multi-interaction vulnerabilities. 
\pocgen~\cite{simsek2025pocgen} outperforms the above approaches. 
However, \pocgen assumes the availability of explicit vulnerability reports, limiting its applicability to newly fixed vulnerabilities. 
\framework complements this by operating in a setting where vulnerability descriptions are unavailable. 

\textbf{Self-Evolving and Prompt Evolution.}
Recent work explores self-evolving LLMs that iteratively improve outputs using feedback~\cite{madaan2023self, shinn2023reflexion, agrawal2026gepa, pryzant2023automatic, yang2024large, gou2024critic, zhao2025pareto}. \textit{Self-Refine}\cite{madaan2023self} enables models to generate, critique, and refine their outputs in multiple iterations, while \textit{Reflexion}\cite{shinn2023reflexion} leverages verbal self-reflection and memory to learn from past failures and improve across trials. 
GEPA~\cite{agrawal2026gepa} is a prompt-evolution technique that uses reflective feedback and evolutionary optimization to improve prompts.
\framework adopts GEPA's Pareto-selection mechanism, but instantiates its evaluation for vulnerability analysis.
Specifically, it refines vulnerability-context dimensions for assessing failed exploit-generation attempts and structures the feedback for evolution.
These dimensions define the domain-specific scores for structuring feedback to guide GEPA-style evolution toward valid PoC exploits.

\textbf{Meta Prompting.}
Meta-prompting techniques allow LLMs to reason about or generate prompts themselves, 
rather than being used directly to solve a task~\cite{hong2024metagpt, khattab2023dspy, suzgun2024meta}. 
Instead of crafting a fixed prompt by hand, the system uses feedback and prior outcomes to produce improved prompts dynamically. 
Similarly, \framework treats failed exploit-generation prompts as evolvable artifacts, and uses failure signals from unsuccessful PoC exploit attempts to synthesize new candidate prompts from iterative refinement of a population of prompts. 

\section{Conclusion and Future Work}
\label{sec:conclusion}
We argue for the need for exploit generators to consider a setting where generating PoC exploits directly from vulnerability-fixing commits before detailed analyses are available.
We present \framework, an automated PoC exploit generator.
To adapt to vulnerability-specific conditions, \framework designs task-specific criteria for assessing vulnerability-related context and uses the resulting feedback to guide prompt evolution.
Our evaluation on SecBench.VFC.js shows that prompt evolution improves both evaluated Exploit Generators across the two underlying LLMs.
With GPT-4o-mini, \framework (LLM) improves direct LLM prompting from 19.5\% to 28.4\%, while the full \framework improves \pocgen from 48.4\% to 58.4\%.
With Qwen3.7-Plus, the corresponding success rates increase from 77.9\% to 85.3\% and from 64.2\% to 79.5\%, respectively.
When vulnerability reports are available, our \framework can achieve a success rate of 71.7\%, improving over \pocgen by 11.1\%.
For the full \framework, the average cost per evaluated vulnerability is \$0.0895 with GPT-4o-mini and \$0.2492 with Qwen3.7-Plus.
We discuss the implications of our findings, including challenges for coordinated disclosure.

In the future, we will extend \framework beyond JavaScript and npm packages to support a broader range of programming languages and software ecosystems.
Future work can also systematically evaluate a wider range of LLMs and Exploit Generator configurations, providing deeper insights into the strengths and limitations of different models and generation strategies.

\section*{Data Availability}
All artifacts, including code, benchmarks are available at: \href{https://github.com/manh-td/pocevolve}{https://github.com/manh-td/pocevolve}

\bibliographystyle{IEEEtran}
\bibliography{reference}

\end{document}